\documentclass[fleqn,twoside]{article}
\usepackage{espcrc2}

\usepackage{graphicx}
\usepackage[figuresright]{rotating}
\usepackage{epsfig}
\usepackage{color}
\newcommand{\AmS}{{\protect\the\textfont2
  A\kern-.1667em\lower.5ex\hbox{M}\kern-.125emS}}

\title{Higgs production at NNLO in QCD: the VBF channel}

\author{P. Bolzoni\address[desy]{Deutsches Elektronen-Synchrotron, DESY, \\
        Platanenallee 6, D15738 Zeuthen, Germany}
        \thanks{Speaker},
        M. Zaro \address[cp3]{Center for Particle Physics and Phenomenology (CP3), \\
        Universit\'e Catholique de Louvain, \\
        B-1348 Louvain-la-Neuve, Belgium},
        F. Maltoni\addressmark[cp3],
        S. Moch\addressmark[desy]}

\begin{document}

\begin{abstract}
We present a phenomenological study for the production of the Higgs
boson at next-to-next-to-leading order (NNLO) in QCD via the vector boson fusion (VBF) process.
After a general discussion about the different production channels
of the Higgs, we show results for hadron colliders like LHC and Tevatron
in VBF. The theoretical predictions are obtained using the structure
function approach. This approximation turns out to be more accurate
than the precision to which the VBF Higgs production channel can
be considered a well defined process by itself and the
theoretical uncertainty are of the order of 1-2\%. The
uncertainties due to parton distributions are also discussed and are
estimated to be at the same level.

\vspace{1pc}
\end{abstract}

\maketitle

\section{INTRODUCTION}

To discover whether the Standard Model Higgs boson exists or not
is one of the aims with most priority at the nowadays
hadron colliders. These are the proton-anti-proton Tevatron
machine at Fermilab and the proton-proton LHC machine at CERN.
To achieve this goal an estimate of the expected events and the
control over those processes which represent the background
noise is required. Nevertheless in the case of a discovery
of the Higgs it would become necessary to also investigate
precisely its properties. On its turn this demands a determination
of the cross section in the various production channels as
precise as possible.

\begin{figure}[htb]
\centering
\includegraphics[scale=0.30,angle=-90]{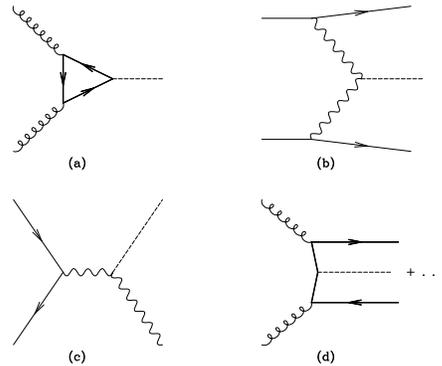}
\vspace*{-15pt}
\caption{Higgs boson production channels at hadron colliders.}
\label{fig:hprodmech}
\end{figure}

At hadron colliders the Higgs production channels which
have a large enough cross section to be relevant
are the gluon-gluon fusion Fig.\ref{fig:hprodmech}(a),
the VBF Fig.\ref{fig:hprodmech}(b) and the
associated production with $W,Z$ bosons and $t\bar{t}$
Fig.\ref{fig:hprodmech}(c,d).

Generally the dominant production mechanism is represented
by the gluon fusion mediated by a top quark loop. For this
production channel the first NNLO corrections in QCD have
been computed in the context of the effective theory in
the limit large top quark mass limit
\cite{Harlander:2000mg,Anastasiou:2002yz,Steinhauser:2002rq,Catani:2003zt}.
It is also known that this approximation works very well up
to the NNLO. Indeed the impact of the finite top mass effects
are about a factor of ten smaller than the uncertainty due
to the scale variation at the same order 
\cite{Harlander:2009my,Marzani:2008az,Pak:2009dg}.
As far as the associated production of the Higgs with
$W,Z$ bosons is concerned, the NNLO QCD corrections have been
implemented in \cite{Brein:2003wg}. Thanks to these results
the theoretical uncertainty is reduced to about 10\% for
the gluon fusion production mechanism and to less than
a few percent for the associated production with gauge bosons.
Very recently NNLO corrections in QCD have been included also
for the VBF production mechanism \cite{Bolzoni:2010xr} via the
structure function approach thus reducing the theoretical
uncertainty for this channel from the 5-10\% of the NLO QCD
and electroweak combined computations
\cite{Han:1992hr,Ciccolini:2007ec}
down to 1-2\%. This makes VBF the theoretically most
accurate Higgs boson production channel at hadron colliders.

\section{THE VBF SIGNAL}

Among the production mechanisms of the Higgs boson that
we have mentioned above, the VBF production mechanism
can be considered by itself as a signal where its
 ``background'' includes also Higgs boson
production by mechanisms other than VBF (see Fig.\ref{fig:hprodmech}).

\begin{figure}[htb]
\centering
\includegraphics[scale=0.33,angle=-90]{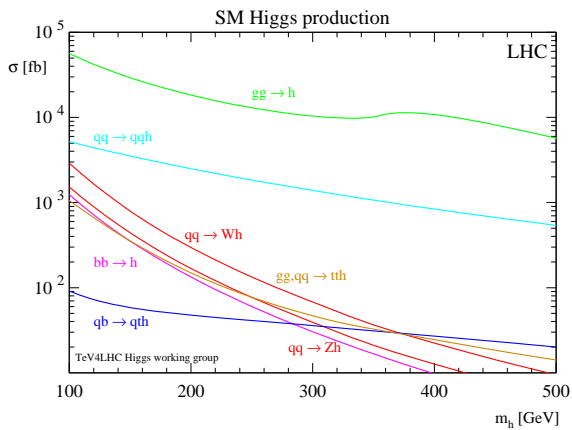}
\vspace*{-15pt}
\caption{Higgs production cross sections at the LHC for the
various channels (see Ref.\cite{Hahn:2006my} for more details about
this plot).}
\label{fig:hcross}
\end{figure}

Here we briefly describe its characteristics.
Due to the preference of the Higgs to couple with heavy particles the VBF at LHC is second
in size only to the gluon fusion channel which is
mediated by a top quark loop (see Fig.\ref{fig:hcross}).
Furthermore it provides a
clean experimental signature: it usually consists of two
almost back to back hard tagging jets (generally with an
invariant mass bigger than $600\, \rm{GeV}$ and a
pseudo-rapidity separation between them $\Delta\eta>4$)
and the Higgs decay is confined in the central rapidity
region. Imposing these constraints to the
invariant mass and the rapidity of the
jets as additional cuts (usually called VBF cuts) one
reaches an impressive improvement of the signal-to-background
ratio \cite{Harlander:2008xn}.

An heuristic way to qualitatively understand
the property of the VBF signature is to consider
angular ordering which arises from the dominant
soft gluon emissions.
One finds that the averaged
azimuthal emission is confined to a cone limited by the
angle between the emitter and the spectator parton.
It can easily be shown that because of the behavior of the
propagators in the $t$-channel
in Fig.\ref{fig:hprodmech}(b)
small values of the emission angle are favored.
Because of this the hadronic activity
in the VBF production mechanism is concentrated in the
forward/backward region with respect to the incoming
colored partons. This is in contrast to
the gluon fusion channel where the radiation is
expected and found in the central rapidity region
\cite{DelDuca:2004wt}.

Keeping in mind that what we strictly call
VBF signal is the Higgs boson weak production with a
color singlet exchange in the $t$-channel, we want now
to discuss the possible interference effects
with other processes.
Already at LO the VBF process in
Fig.\ref{fig:hprodmech}(b) can interfere with the
production mechanism in Fig.\ref{fig:hprodmech}(c)
where the associated produced gauge boson decays into
two quarks. However as it is shown in \cite{Ciccolini:2007ec}
this interference effect is at the per mil level.
At higher orders, interference effects occur with the
gluon fusion production mechanism in Fig.\ref{fig:hprodmech}.
Also in this case the interference effect is small
\cite{Andersen:2006ag,Andersen:2007mp} and it
turns out to be well below the percent level.
This demonstrates that the Higgs boson production via the VBF
mechanism can be defined a signal by itself within an
ambiguity better than the 1\%. It is this ambiguity
in defining the Higgs boson production via VBF that
sets also the target theoretical precision for this
observable.

\section{THE COMPUTATION}

Here we want to illustrate the structure function approach
in the VBF production chnnel \cite{Han:1992hr} and show that
this approximation remains sufficiently accurate even at NNLO
in QCD.

\begin{figure}[htb]
\vspace{5pt}
\centering
\includegraphics[scale=0.475]{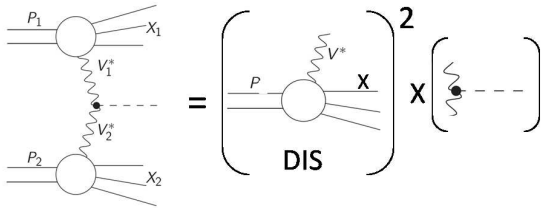}
\vspace*{-15pt}
\caption{Schematic view of the structure function approach.}
\label{fig:strucfuncapp}
\end{figure}

The structure function approach consists basically in
viewing the VBF process as a double deep-inelastic scattering
(DIS) attached to the colorless pure electroweak vector boson
fusion into a Higgs boson. A qualitative illustration of the
structure function approach is given in Fig.\ref{fig:strucfuncapp}.
According to this approach one can include NLO QCD corrections
to the VBF process employing the standard DIS structure functions
$F_i(x,Q^2);\,i=1,2,3$ at NLO \cite{Bardeen:1978yd}.
Similarly at the NNLO level one has to employ the
corresponding structure functions
\cite{Kazakov:1990fu,Zijlstra:1992kj,Zijlstra:1992qd,Moch:1999eb}.

The structure function approach represents a very
accurate approximation because it is
based on the absence or smallness of the QCD interference
contributions between the two inclusive final states $X_1$
and $X_2$. We now discuss the various
contributions up to NNLO which in principle violate the structure
function approach but which can nevertheless be
safely neglected.

At LO there is already a structure function violating
contribution coming from the interferences between identical
final state quarks (e.g. $uu\rightarrow Huu$) or between
processes where either a $W$ or a $Z$ can be exchanged
(e.g. $ud\rightarrow Hud$). Simple kinematical arguments
show that such contributions are very small and contribute
to the total cross section well below the percent
level \cite{Dicus:1985zg}.
These contributions can be easily computed and have been
included in our results anyway.

At the NLO level possible contributions violating
the structure function approach arise when a gluon in the $t$-channel
is exchanged between the upper and the lower quark line
in Fig.\ref{fig:hcross}(b).
However the interference of such one-loop contributions
with the LO diagram have a vanishing color factor due to the
generators $t^a$ $(a=1,\dots,N_c^2-1)$ of the color group $SU(N_c)$
being traceless. This means that apart from the interference
effects discussed at LO the structure function
approach represents an exact approach to the computation.

\begin{figure}[htb]
\centering
\includegraphics[scale=0.45]{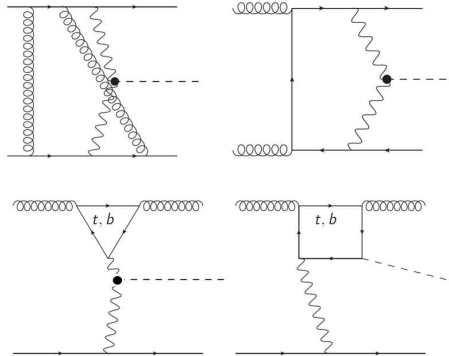}
\vspace*{-15pt}
\caption{Examples of neglected Feynman diagrams at NNLO.}
\label{fig:neglect}
\end{figure}

At NNLO the structure function approach is not exact but
it can be still considered a very good approximation.
The types of
diagrams that violate the structure function approach
are shown in Fig.\ref{fig:neglect}. The first type represents
a double gluon-exchange in the $t$-channel (note that one
or both of the two emissions could also be real);  the second type is
a representative of the so-called single quark line (SQL)
diagrams contributing at NNLO; the last two are heavy quarks
(top and bottom) loop diagrams. The first type of contributions
represents a gauge invariant, infrared and ultraviolet finite
class of diagrams. Another characteristic of this class of diagrams
is its color suppression by a factor of $1/N_c^2$ with
respect to the contributions included by the structure
function approach. Furthermore this type of contributions
are also strongly kinematically suppressed \cite{vanNeerven:1984ak,Blumlein:1992eh,Figy:2007kv}.
This is mainly due to the behavior of the gluon propagator
in the $t$-channel and/or to the small overlapping of the phase
space of  real emissions from the upper quark line and real emissions from the lower one.
The neglected SQL type contributions in Fig.\ref{fig:neglect} do not represent
a class of infrared safe diagrams. However as shown in
\cite{Harlander:2008xn} their impact is
small enough not to produce a significant deterioration
of the VBF signal. Also, these color exchange effects are,
by our definition, no VBF processes.
Finally we take into consideration the triangle and the
box contributions in Fig.\ref{fig:neglect}.
\begin{table*}[htb]
\newcommand{\m}{\hphantom{$-$}}
\newcommand{\cc}[1]{\multicolumn{1}{c}{#1}}
\renewcommand{\tabcolsep}{2pc} % enlarge column spacing
\renewcommand{\arraystretch}{1.2} % enlarge line spacing
\begin{tabular}{@{}llll}
\hline
$m_h$ (GeV)\           & \cc{$120$} & \cc{$300$} & \cc{$500$} \\
\hline
$1.96$  TeV               & \m3.87\rm\,E-6 (0.0690) & \m2.52\rm\,E-7 (0.0054)& \m1.50\rm\,E-8  (0.00042) \\
 $7$ TeV    & \m2.62\rm\,E-4 (1.235) & \m7.89\rm\,E-5 (0.614)& \m2.73\rm\,E-5 (0.088) \\
\hline
\end{tabular}\\[2pt]
\caption{\label{table:triangle}
Total cross section (pb) from the neglected triangle
diagram in Fig.\ref{fig:neglect}. In parenthesis are reported
also the numbers included in the NNLO computation in the structure
function approach employing the MSTW PDF set \cite{Martin:2009iq}.
All the parameters have been taken from PDG
2008 \cite{Amsler:2008zzb}.}
\end{table*}
Even if a full computation is in progress \cite{InProgress:2010}
as a first rough estimation we have computed
the triangle contribution in
Fig.\ref{fig:neglect} in the limit of infinite top
mass. In Table~\ref{table:triangle} we report some values of the
contribution to the total cross section from the triangle
at $1.96\,\rm{TeV}$ for the Tevatron and at $7\,\rm{TeV}$
for the LHC. This has been checked performing
two independent computations. As we can see from
Table \ref{table:triangle} its impact is very small and
can be safely neglected.

\section{RESULTS AT HADRON COLLIDERS}

Here we discuss some numerical results obtained for the
VBF production mechanism at hadron colliders up to the
NNLO in QCD employing the structure function approach.

\begin{figure}[htb]
\centering
\includegraphics[scale=0.55]{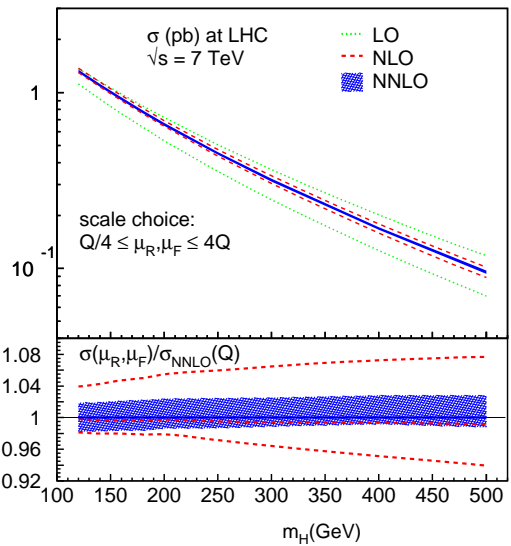}
\vspace*{-15pt}
\caption{Total cross section for the LHC at
$7\,\rm{TeV}$ employing the MSTW PDF set \cite{Martin:2009iq}.}
\label{fig:plot_7_sc1}
\end{figure}
\begin{figure}[htb]
\centering
\includegraphics[scale=0.55]{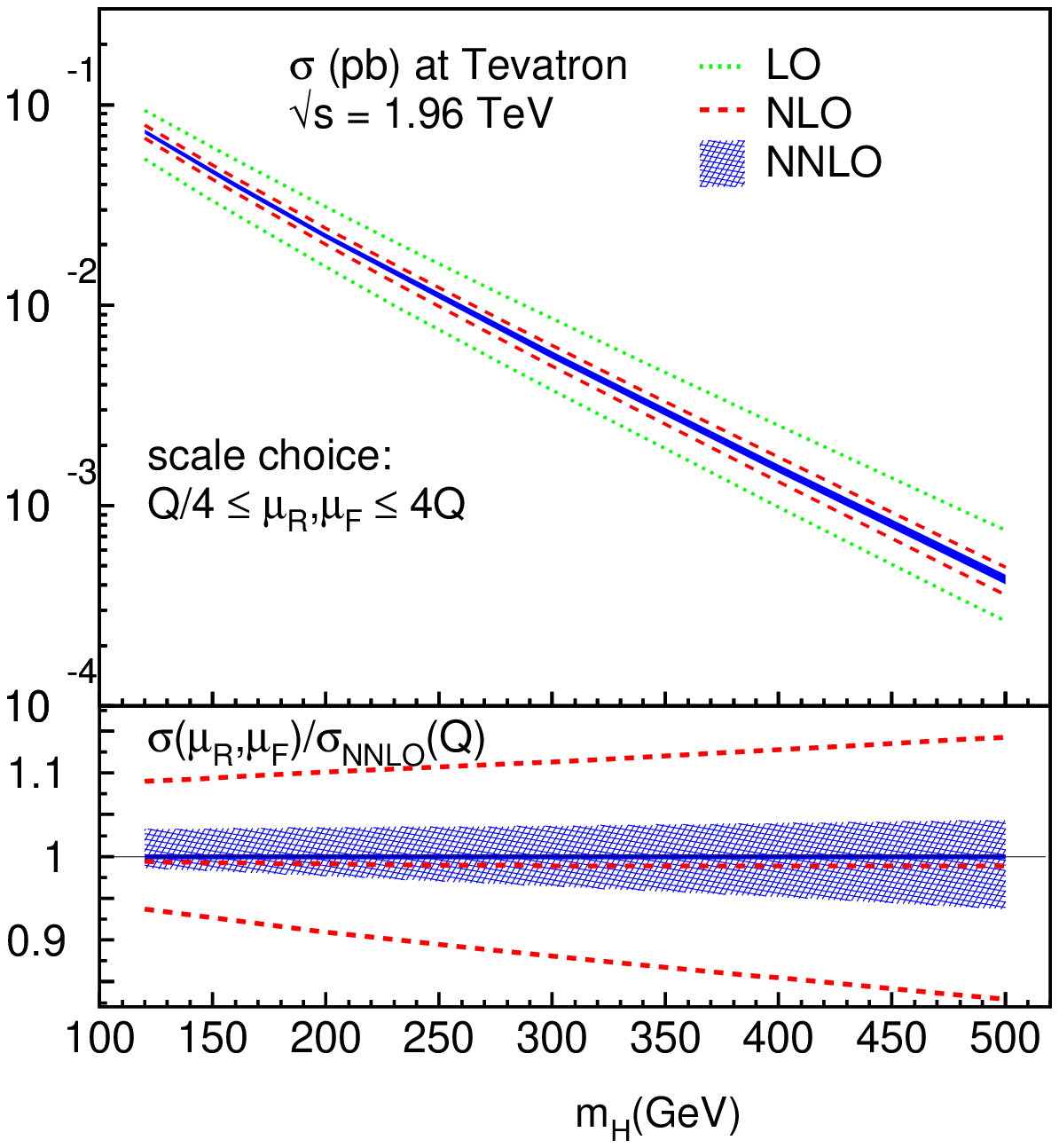}
\vspace*{-15pt}
\caption{The same as Fig.\ref{fig:plot_7_sc1} for the
Tevatron.}
\label{fig:plot_TEV_sc1}
\end{figure}
\begin{figure}[htb]
\centering
\includegraphics[scale=0.375]{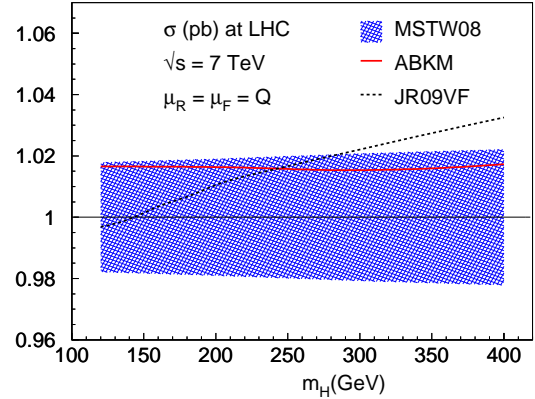}
\vspace*{-15pt}
\caption{The PDF uncertainty of the total cross
section at NNLO for LHC at $7\,\rm{TeV}$ for the 68\% CL MSTW
PDF set \cite{Martin:2009iq}. For ABKM \cite{Alekhin:2009ni}
and JR09VF \cite{JimenezDelgado:2009tv} the ration of the
central value is plotted.}
\label{fig:plot_7_sc1_pdf}
\end{figure}
\begin{figure}[htb]
\centering
\includegraphics[scale=0.375]{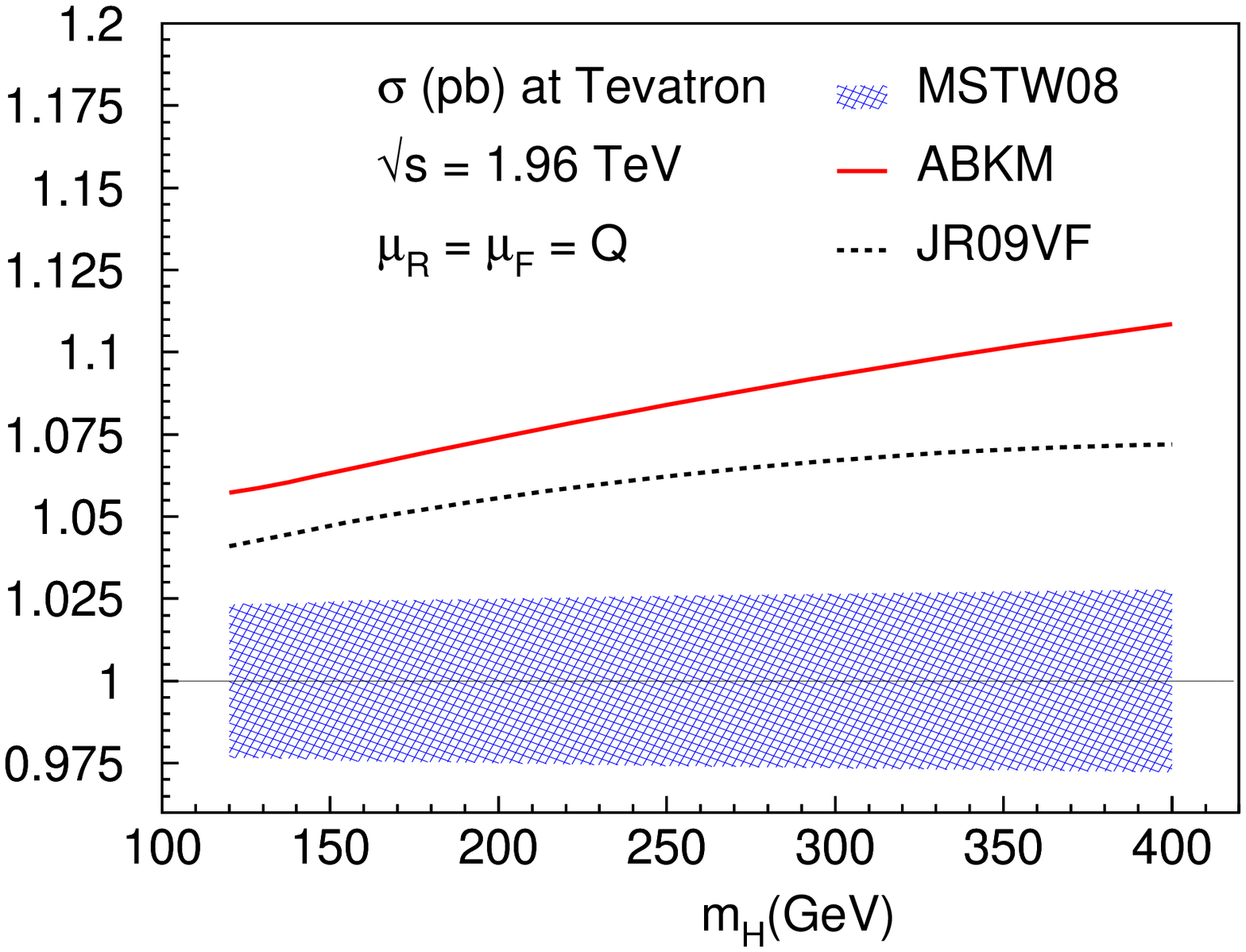}
\vspace*{-15pt}
\caption{Same as Fig.\ref{fig:plot_7_sc1_pdf}
for the Tevatron.}
\label{fig:plot_TEV_sc1_pdf}
\end{figure}

For the electroweak parameters we have used the central values
released by the PDG collaboration in 2008 \cite{Amsler:2008zzb}
while for the parton distributions functions we choose the
set MSTW \cite{Martin:2009iq}. For the LHC we set the center
of mass energy to be $\sqrt{S}=7\,\rm{TeV}$ for the LHC and
$\sqrt{S}=1.96\,\rm{TeV}$ for the Tevatron.

In Fig.\ref{fig:plot_7_sc1} we plot the total cross section
for the VBF production mechanism at the LHC at $7\,\rm{TeV}$.
The LO, NLO and NNLO results in QCD are shown
as a function of the Higgs boson mass which first of all
induce only a rather mildly dependence on it. The bands represent
the theoretical uncertainty of the prediction. They have
been obtained varying the factorization and the
renormalization scales in the quite large range
$\mu_{\rm{R(F)}}\in [Q/4,4Q]$ where $Q$ is the virtuality
of the vector bosons which ``fuse'' into the Higgs.
Clearly other scale choices are possible (e.g. the
choice $Q=m_h$ the Higgs mass) but the one chosen for
the plot turned out to be the more natural choice because
it exhibits a better convergence of the perturbative
expansion. This also shows that at NNLO
in QCD the theoretical uncertainty is reduced to be less than the
$2\%$ reaching the same level of ambiguity at which
the Higgs production signal via VBF can be
defined phenomenologically.

In Fig.\ref{fig:plot_TEV_sc1} we report on numbers
for the Tevatron where the center of mass energy
is set to $1.96\,\rm{TeV}$. As one can expect the total
cross section shows the same behavior upon varying the
Higgs mass and it is almost an order of magnitude
smaller. Again the lower part of Fig.\ref{fig:plot_7_sc1}
shows a very good convergence of the perturbative
QCD expansion. Even if the theoretical uncertainty remains
slightly bigger compared to the LHC the relative
improvement with respect to the NLO prediction is at the
same order of percentage.

Finally we consider also the uncertainties coming
from the parton distributions. To achieve this we have
employed the MSTW $68\%$ confidence level PDF sets
\cite{Martin:2009iq} and compare with other NNLO
PDF sets, i.e. ABKM \cite{Alekhin:2009ni} and JR09VF
\cite{JimenezDelgado:2009tv}. The results in
Figs.\ref{fig:plot_7_sc1_pdf},
\ref{fig:plot_TEV_sc1_pdf} show that
an almost constant $2\%$ PDF uncertainty can be associated
to the cross section for both the LHC and the Tevatron.
In the case of the Tevatron the difference between the
MSTW, ABKM and JR sets is due to larger uncertainties
for the high-$x$ quark PDFs \cite {Alekhin:2009ni}.

To conclude we give the address of the web interface
where our code for the NNLO VBF total cross section
can be used online \cite{WebInterface:2010}. After
the registration, setting on a dialog window the energy and choosing
the hadron collider the numerical answer is received per email.

\section{CONCLUSIONS}

We have described recent progress for the VBF Higgs predictions
and investigated at which level it can be considered a well
defined process by itself.
Then after showing how well the structure function approach works even
at the NNLO in QCD we have employed it to obtain
predictions at the hadron colliders. The theoretical
uncertainty is estimated to be less than $2\%$ which is compatible
with the ambiguity at which this signal can be defined.
Finally what has been presented here is a natural first step towards
less inclusive (e.g. rapidity distributions) predictions
of this process at NNLO in QCD.

\bibliographystyle{h-physrev5}
\bibliography{bibvbf}

\end{document}